\documentclass[prl,twocolumn,showpacs,superscriptaddress]{revtex4-1}
\usepackage{graphicx,amssymb,amsmath,psfrag,color}
\begin{document}
\definecolor{darkgreen}{rgb}{0,0.5,0}
\definecolor{gold}{rgb}{0.549,0.549,0}

\title{Quantum quench in 2D using the variational Baeriswyl wavefunction}

\author{Bal\'azs D\'ora}
\email{dora@eik.bme.hu}
\affiliation{Department of Physics and BME-MTA Exotic  Quantum  Phases Research Group, Budapest University of Technology and
  Economics, 1521 Budapest, Hungary}
\author{Masudul Haque}
\affiliation{Max-Planck-Institut f\"ur Physik komplexer Systeme, 01187 Dresden, Germany}
\author{Frank Pollmann}
\affiliation{Max-Planck-Institut f\"ur Physik komplexer Systeme, 01187 Dresden, Germany}
\author{Bal\'azs Het\'enyi}
\affiliation{Department  of  Physics,  Bilkent  University  06800,  Ankara,  Turkey}

\date{\today}

\begin{abstract}

By combining the Baeriswyl wavefunction with 
equilibrium and time-dependent variational principles, we develop a non-equilibrium formalism to
study quantum quenches for two dimensional spinless fermions with
nearest-neighbour hopping and repulsion.  The variational ground state energy
and the short time dynamics agree convincingly with the results of numerically
exact simulations.  We find that depending on the initial and final
interaction strength, the quenched system either exhibits undamped
oscillations or relaxes to a time independent steady state.
The time averaged expectation value of the CDW order parameter rises sharply
when crossing from the steady state regime to the oscillating regime,
indicating that the system, being non-integrable, shows signs of
thermalization with an effective temperature above or below the equilibrium
critical temperature, respectively.
\end{abstract}


\maketitle

\paragraph{Introduction ---}

The past decade has witnessed a massive resurgence of interest in the coherent
dynamics of quantum many-body systems far from equilibrium 
\cite{dziarmagareview,polkovnikovrmp,eisertreview_Nature2015}.  This activity
has been catalyzed mostly by developments in cold atom experiments 
\cite{BlochDalibardZwerger_RMP08,polkovnikovrmp}, which allow for long
coherence and explicitly tracking the evolution of well-isolated many-body
systems in real time.  At the same time, there is also a revival of interest
in the ultrafast time evolution of solid-state systems.  A ubiquitous paradigm
in the theoretical study of non-equilibrium dynamics is the so-called quantum
quench, where one starts from the ground state of a Hamiltonian, and then
instantaneously changes a parameter, so that the system wavefunction evolves
under a different Hamiltonian.


In contrast to the one-dimensional (1D) case, there are only few exact theoretical methods
available to treat real-time evolution in 2D quantum systems.  For example, the DMRG
technique 
\cite{White-1992,Kjaell-2012,dmrgmps,manmana05,zangara13,daley04,zaletel14} is
challenging in 2D, there are almost no 2D situations that are integrable via
the Bethe ansatz, and exact numerical diagonalization is faced with a severely
rapid growth of Hilbert space size.  There are thus relatively few studies of
interaction quenches in 2D, e.g., Refs. \cite{hamerla,hauschild} focusing mainly on
short time dynamics, or Ref. \cite{goth} on time evolution in the
absence of interactions.


Variational wavefunctions are vital in equilibrium condensed matter physics,
e.g., the variational BCS wavefunction for superconductivity, the Yosida
wavefunction for the Kondo model~\cite{Yosida1966}, and the Gutzwiller wavefunction
(GWF)~\cite{Gutzwiller63,Gutzwiller65} for the Hubbard model.  The parameters
of a well-chosen variational wavefunction can incorporate the most essential
physics of a complex many-body system.  In contrast to equilibrium, the
non-equilibrium variational principle~\cite{kramer} has been used less
extensively in correlated systems.  The bosonic version of the GWF was often
used during the last decade to explore dynamics in the Bose-Hubbard model~ 
\cite{RokhsarKotliar, JakschCiracZoller_PRL2002,Jreissaty}.  Recently,
computational methods using Monte Carlo evaluation were formulated for the
evolution of variational wavefunctions with large numbers of
parameters~\cite{BeccaSchiroFabrizio_SciRep2012, becca_PRA2014, Imada_arXiv1507}.  In 
addition, the GWF within the Gutzwiller
approximation~\cite{Gutzwiller63,Gutzwiller65} was recently used to study the
time evolution~\cite{schiro} of the fermionic Hubbard model.  This
approximation gives the exact solution of the GWF in infinite
dimensions~\cite{Metzner89}.


The GWF~\cite{Gutzwiller63,Gutzwiller65} consists of a
non-interacting wavefunction acted on by an operator which
projects out states not favoured by the interaction (i.e. double occupation
for the spinful Hubbard model).  A completely opposite variational approach, the Baeriswyl
wavefunction (BWF)~\cite{Baeriswyl86,valenzuela}, is based on a fully
localized wavefunction (an eigenfunction of the interaction term) and is acted
on by a kinetic energy projector.  The projector has the effect of promoting
the hopping of particles in an otherwise completely localized system.  In both
of these approaches, the strength (coefficient) of the projection operator
serves as the variational parameter.

In the present work, we will use the BWF to describe a model of 2D spinless fermions on the square
lattice with nearest neighbor repulsion at half filling.  The reference state is a charge density
wave (CDW) with one of the sublattices exactly filled.  An advantage of this approach, as we show,
is that it is possible to perform the calculations exactly within the variational manifold, without
the requirement of additional approximations.  In addition, the wavefunction is \emph{exact} at the
two extreme limits of infinite and zero interaction.  We show that equilibrium properties are
well-reproduced at all interactions. We use the time-dependent variational principle\cite{kramer,seibold} to investigate
the dynamics after interaction quenches, and show that the system either relaxes to a steady state
or oscillates indefinitely.  We map out the ``quantum quench phase diagram'' according to this
criterion and conjecture an interpretation based on the thermalization temperature.  The analysis
involves characteristics of the CDW order parameter.  Density waves have long been a central topic
in the study of electronic phases in the solid state~\cite{gruner}.  Recent experiments have also
addressed their non-equilibrium dynamics for CDW~\cite{trotzky} as well as of related spin density
(antiferromagnetic) patterns~\cite{PhilipsPorto_PRL2007, Trotzky_Science2008}.  The present work
studies CDW dynamics for a 2D interacting system, a topic which at present lies outside the reach of
exact methods.

\paragraph{Model ---}

\begin{figure}[h!]
\centering
\includegraphics[width=7cm]{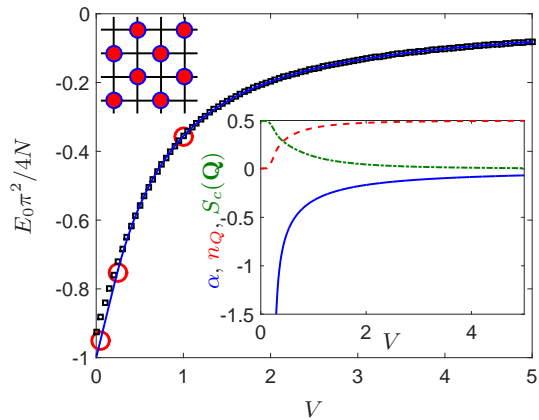}
\caption{(Color online) The variational ground state energy, normalized by 
its non-interacting  value (blue solid line), is compared to 2D DMRG data (red circles) and ED on $4\times 4$ cluster (black squares).
The lower inset visualizes the 
CDW order parameter $n_Q$ (red dashed line), the equal time structure factor from Eq.~
\eqref{strucfac} for momentum $\bf Q$ (green dash-dotted line)
and the variational parameter, $\alpha$ (blue line) as a
function of the interaction strength. The upper inset depicts a small segment of  the fully polarized CDW state.}
\label{gsenergy}
\end{figure}

The spinless fermionic Hubbard model is defined on the 2D square lattice as $H=H_{kin}+H_{int}$ with
\begin{gather}
H_{kin}=-\frac J2\sum_{\langle n,m\rangle}c^+_nc_m, \hspace*{2mm} H_{int}=\frac V2\sum_{\langle n,m\rangle}n_nn_m,
\label{hamilton}
\end{gather}
where $n_n=c^+_nc_n$ is the particle density operator at site $n$, and $c_n$
annihilates a fermion from site $n$.  The sum over $\langle n,m\rangle$ runs through all lattice sites and keeps only nearest-neighbor pairs.  $J$ is the
strength of the single-particle hopping integral, used as the unit of energy
in the following, thus it is suppressed, $V>0$ is the strength of the nearest
neighbor Coulomb repulsion. The BWF amounts to starting from
the $V\rightarrow\infty$ state and using the kinetic energy as a projection
operator~\cite{valenzuela}
\begin{gather}
|\Psi_B\rangle={N_B^{-1}}\exp\left(\tilde\alpha H_{kin}\right)|\textmd{CDW}\rangle,
\label{bwf}
\end{gather}
where $|\textmd{CDW}\rangle$ is the charge ordered ground state $H$ 
in the atomic limit, where the charges follow a checkerboard pattern on
the lattice sites (see Fig. \ref{gsenergy}), $N_B$ is the overall
normalization factor.

The BWF, due to its CDW ground state~\cite{gruner} possesses an appealing wavefunction in momentum space as well.
The fully polarized CDW wavefunction is
$|\textmd{CDW}\rangle=\prod\limits_{{\bf  k}\in \textmd{RBZ}}\frac{1}{\sqrt {2}}\left(c^+_{\bf k}+c^+_{\bf k-Q}\right)|0\rangle$,
where ${\bf Q}=(\pi/a,\pi/a)$ is the ordering wavevector, $a$ the lattice constant,
the product goes through the reduced (magnetic) Brillouin zone (RBZ) and $|0\rangle$ is the fermionic vacuum.
The normalized variational wavefunction assumes the form
\begin{gather}
|\Psi_B\rangle=\prod\limits_{{\bf k}\in \textmd{RBZ}}\frac{\exp[\tilde\alpha\epsilon({\bf k})]c^+_{\bf k}+\exp[-\tilde\alpha\epsilon({\bf k})]c^+_{\bf k-Q}}{\sqrt{2\cosh[2\textmd{Re}\tilde\alpha\epsilon({\bf k})]}}|0\rangle
\label{bwf1}
\end{gather}
with $\epsilon({\bf k})=-\cos(k_xa)-\cos(k_ya)$.  Here,
$\tilde\alpha=\alpha+i\eta$ is the complex variational parameter.  Its
imaginary part becomes relevant when studying the quench problem.

\paragraph{Equilibrium properties ---}
 The variational problem is solved by minimizing the ground state energy with respect to the variational parameter $\tilde\alpha$
as
$E_0=\min\limits_{\tilde\alpha} \langle \Psi_B|H|\Psi_B\rangle$,
and our conventions imply that negative real values of $\tilde\alpha$ minimize the above functional, namely $\alpha<0$ and $\eta=0$.
The expectation value of the kinetic and interaction energy is evaluated \emph{exactly} with the variational wavefunction as
$\langle H_{kin}\rangle=\epsilon({\bf 0})I_2$ and 
\begin{gather}
\langle H_{int}\rangle=-NV\frac{\epsilon({\bf 0})}{4}+\frac{V \epsilon({\bf 0})}{N} \sum_{i=1,2,3} I_i^2,\label{eint}
\end{gather}
where $N$ is the total number of lattice sites and~\cite{epaps}
\begin{subequations}
\begin{gather}
I_1=\sum_{\bf k}\frac{\cos[2\eta\epsilon({\bf k})]}{2\cosh[2\alpha\epsilon({\bf k})]},\\
I_2=\sum_{\bf k}\frac{\cos(k_xa)}{2}\tanh[2\alpha\epsilon({\bf k})],\\
I_3=\sum_{\bf k}\frac{\cos(k_xa)\sin[2\eta\epsilon({\bf k})]}{2\cosh[2\alpha\epsilon({\bf k})]}.
\end{gather}
\end{subequations}
Note that we obtain an exact, closed expression for the variational energy, $\langle H_{kin}\rangle+\langle H_{int}\rangle$ in
2D for this model.  So far, closed expressions have not been derived
based on the GWF for any two-dimensional model.  Moreover, the energy expectation
values are valid also
for hypercubic lattices in arbitrary dimension $d$, after making the
replacement $\epsilon({\bf k})=-\sum_{i=1}^d\cos(k_ia)$, and the $\bf k$ sums
run over $d$ wavevector components.

\begin{figure}[h!]
\centering
\includegraphics[width=7cm]{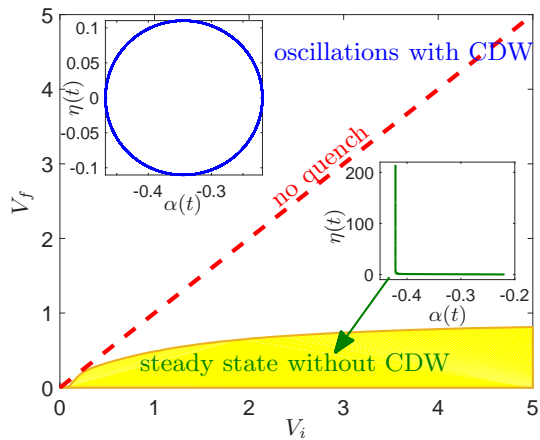}
\caption{(Color online) The quantum quench phase diagram for $V_i$ and $V_f$ is shown. For small $V_f$, a time independent steady state is reached which
could be interpreted as an indicator thermalization at an effective temperature higher than the equilibrium CDW transition temperature
with $\eta(t\rightarrow\infty)\sim t$ and
$\alpha(t)$ saturating to a finite value, as depicted in the lower inset for $V_i=1.5$, $V_f=0.25$ and $0<t<200$.
For sizable quenches, on the other hand, the system periodically returns to its initial state, visualized in the upper inset for $V_i=1.5$, $V_f=1$.
For $V_i=\infty$, the transition from the steady state to oscillating behaviour occurs at $V_f\approx 0.97$.
\label{phase2d}}
\end{figure}

The occupation number of momentum states is $n_{\bf k}=\langle c^+_{\bf
  k}c_{\bf k}\rangle=1/\{1+\exp[-4\alpha\epsilon({\bf k})]\}$.  This
expression corresponds to the occupation number of a non-interacting system at
finite temperature with $T = -1/4\alpha$.  In the following we will make use
of this interpretation.  The initial CDW state thus possesses infinite
temperature with $n_{\bf k}=1/2$, as follows from $|\textmd{CDW}\rangle$ as
well, albeit it differs from a trivial $T=\infty$ state due to the anomalous
correlations with wavevector $\bf Q$ from the CDW.  The charge density at
lattice site $\bf R$ is
\begin{gather}
n({\bf R})=\frac 12+n_Q\cos({\bf QR}),\hspace*{5mm}
n_Q=\frac{I_1}{N},
\label{cdwm}
\end{gather}
 with $n_Q$ the  CDW order parameter, describing the $\bf Q$ periodic charge oscillations.

The $\bf Q$th Fourier component of the equal-time connected structure factor
(SF), which measures non-local properties and  CDW correlations, is
\begin{gather}
S_c({\bf Q})=\langle\Psi_B| \rho_{\bf Q} \rho_{\bf -Q}|\Psi_B\rangle-N|n_Q|^2=\nonumber\\
=\frac 12-\frac{1}{2N}\sum_{\bf k}\frac{\cos^2[2\eta\epsilon({\bf k})]}{\cosh^2[2\alpha\epsilon({\bf k})]},
\label{strucfac}
\end{gather}
where $\rho_{\bf Q}=\frac{1}{\sqrt{N}}\sum_{\bf k}c^+_{\bf k}c_{\bf k-Q}$, and gives $0$ in the fully polarized CDW state and $1/2$ in the absence of CDW correlations,
corresponding to infinite and zero effective temperatures. Therefore, the value of the SF is related to the amount of CDW order and its effective temperature of the  system.
It is time independent in equilibrium, shown in Fig. \ref{gsenergy}, and reflects the behaviour of $n_Q$.

The variational wavefunction is exact in two opposite limits, $V=0$ ($\alpha=-\infty$) and $V=\infty$ ($\alpha=0$). In the former case, 
the ground state energy is $E_0=-4N/\pi^2$, while for the latter, the kinetic energy is suppressed by the CDW and the interaction energy 
reaches its minimum, therefore $E_0=0$.
In between these two extrema,  the ground state energy, the CDW order parameter $n_Q$ 
and the optimal variational parameter are shown in Fig. \ref{gsenergy},
together with numerical data using 2D DMRG and exact diagonalization (ED) on $4\times 4$ cluster.
The 2D DMRG data is obtained by extrapolating the energies for infinitely long cylinders with circumferences $L=6,8,10$ to the thermodynamic limit. 
The agreement between the variational and numerically obtained energies is remarkable. The ED matches the DMRG data for larger $V$, indicating short correlation length, such that the relatively small system size does not
influence the ground state energy.

 The system is always in a CDW state, except for $\alpha\rightarrow -\infty$, corresponding to 
the non-interacting limit $V=0$. This is in agreement with what is expected for the spinless Hubbard model on the square lattice~\cite{foster,dewoul}. 
The CDW phase appears  because of the square shaped Fermi surface at half filling,
in the case of perfect nesting~\cite{gruner}, and also due to the  log-divergent density of states upon approaching  half filling as $\sim\ln(1/\omega)$.

\paragraph{Quantum quench ---} 
We now turn to the investigation of the quantum quench, when an initial $V_i$ is changed suddenly to $V_f$ at $t=0$. The
time-dependent wavefunction is of Baeriswyl form, $|\Psi_B(t)\rangle$ as well, and the quench amounts to allowing 
for time-dependent variational parameters $\alpha(t)$ and $\eta(t)$ in Eq.~\eqref{bwf1}.
Their time-dependence follows from the time-dependent variational principle, which requires the minimization of the real Lagrangian, defined as~\cite{kramer,seibold}
\begin{gather}
L(t)=-\textmd{Im}\left(\langle \Psi_B|\partial_t\Psi_B\rangle\right)-\langle \Psi_B|H|\Psi_B\rangle,
\end{gather}
with respect to the time-dependent variational parameters. 
 Using 
$L(t)= \langle H_{kin}\rangle\left(\partial_t\eta -1\right)-\langle H_{int}\rangle$
with the expectation values taken from Eq.~\eqref{eint} with the time-dependent variational parameters,
we finally arrive to the Euler-Lagrange equations, determining the quench dynamics as
\begin{gather}
\partial_t\alpha=\dfrac{\dfrac{\partial \langle H_{int}\rangle}{\partial\eta}}{\dfrac{\partial\langle H_{kin}\rangle}{\partial\alpha}},\hspace*{5mm}
\partial_t\eta=1+\dfrac{\dfrac{\partial \langle H_{int}\rangle}{\partial\alpha}}{\dfrac{\partial\langle H_{kin}\rangle}{\partial\alpha}},
\label{vareqs}
\end{gather}
supplemented with the initial conditions $\alpha(0)=\alpha_i$, originating from the equilibrium configuration for $V_i$, and $\eta(0)=0$.
From these, the total energy is conserved after the quench, as expected.

\begin{figure}[t!]
\centering  
\includegraphics[width=8cm]{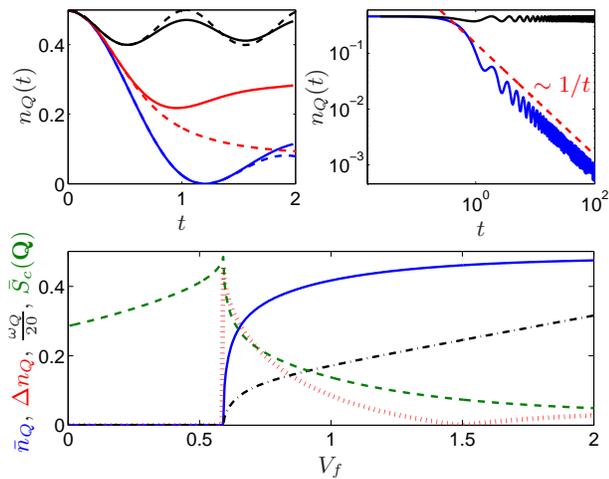}
\caption{(Color online) Top left: The short time decay of the CDW order parameter from time-dependent 2D DMRG for an infinite cylinder of circumference $L=8$ (solid lines) and variational calculation (dashed lines) for $V_i=\infty$ and $V_f=0$ (blue), 1 (red) and 2 (black). Top right: 
The typical 
behaviour of the CDW order parameter is plotted in the steady state region (blue line), vanishing as $1/t$, for 
$V_i=1.5$ and $V_f=0.25$ and  in the oscillatory region (black line) for $V_i=1.5$ and $V_f=1$. 
Bottom: The time average value of the CDW order parameter (blue solid line), the amplitude (red dotted line)
and the frequency (black dash-dotted line) of oscillations, together with the structure factor for momentum $\bf Q$ (green dashed line) are shown for $V_i=1.5$.}
\label{densosc2d}
\end{figure}

From Eq.~\eqref{vareqs}, the two limiting cases are recovered. First, 
when no quench was performed, $\alpha(t)=\alpha_i$ and $\eta(t)=0$.
Second, in the case of quenching from $V_i=\infty$ to $V_f=0$, the $\alpha_i=0$ initial condition fixes $\alpha(t)=0$, while $\eta(t)=t$.
For general values of initial and final interactions, one needs to integrate these coupled differential equations numerically~\cite{epaps}. The obtained phase diagram
is shown in Fig. \ref{phase2d}. 
Eq.~\eqref{hamilton} in 2D is non-integrable, 
therefore it is expected to thermalize after a sudden quench, and we argue that our simple variational scheme is able to capture some of its 
physics.

The time dependence of the CDW order parameter is calculated from
Eq.~\eqref{cdwm} after inserting the time-dependent variational parameters into
$I_1$.  The short time behaviour from the BWF agrees well with that from 2D
time-dependent DMRG obtained using the algorithm introduced in
Ref.~\cite{zaletel14}, as shown in Fig. \ref{densosc2d}. The $V_i=\infty$
$\rightarrow$ $V_f=0$ quench is exact within the variational framework, which
also allows us to check the temporal validity of the numerics.

For longer times, the variational solution either oscillates around its time
average, or reaches a steady state after the quench~\cite{epaps}.  
For quenches with small
$V_f$, a time independent steady state is reached with $\eta(t)$ increasing
linearly with time and $\alpha(t)$ saturating to a fixed value.  This can be
thought of as the heating up of the system, such that it eventually
thermalizes to a higher effective temperature than the equilibrium CDW
transition temperature, thus CDW is absent.  The CDW order parameter decays to
zero since due to $\eta(t)\sim t$, the numerator in $I_1$ oscillates fast and
kills the integral with increasing $t$.  In particular, for arbitrary spatial
dimension $d$, a $n_Q(t)\sim t^{-d/2}$ decay is found, in accord with special
quenches of the 1D Heisenberg chain~\cite{barmettler}.  A similar decay is
found for the antiferromagnetic order of the 2D spinful Hubbard
model~\cite{goth}, quenched to a non-interacting system.  While in 1D, the CDW
order parameter oscillates and changes also sign after the
quench~\cite{barmettler,goth}, it only oscillates around its envelope function
in 2D but does not change sign, as shown in Fig. \ref{densosc2d}.  This is
supported by the exact result when quenching from $V_i=\infty$ to $V_f=0$,
$n_Q(t)=J_0^d(2t)/2$, valid in arbitrary dimension $d$, and $J_0(x)$ is the
zeroth Bessel function of the first kind, in agreement with our numerical
findings.

For larger $V_f$, on the other hand, an oscillating solution is found for both
$\alpha(t)$ and $\eta(t)$, and the system periodically returns to its initial
state, and seemingly does to reach a time independent steady state,
illustrated in Fig. \ref{phase2d}. Within our approach, we interpret this as
an indicator of thermalization to a thermal state with lower effective
temperature than the equilibrium CDW transition temperature, hence the
resulting state possesses a finite CDW order.  The oscillations are artifacts
of the variational calculations, but the value it oscillates around could
correspond to that in the thermalized state.  Throughout the time evolution,
$\alpha(t)$ stays negative, and its equilibrium interpretation as an effective
temperature through the momentum distribution remains true after the quench,
with a time-dependent effective temperature $T(t)=-1/4\alpha(t)$.

In order to further characterize the CDW phase,
we consider the  time average of the order parameter as $\bar n_Q=\lim_{t\rightarrow\infty}\frac 1t \int_0^t n_Q(t^\prime)dt^\prime$, together with the
amplitude, $\Delta n_Q$ and frequency, $\omega_Q$ of the oscillations of $n_Q(t)$ in the oscillating regime, shown in Fig.~\ref{densosc2d} for the representative case of $V_i=1.5$.
Here, $\Delta n_Q=\lim_{t\rightarrow\infty}\left\{\max[n_Q(t)]-\min[n_Q(t)]\right\}$ 
in the oscillating regime and $\omega_Q$ is the basic harmonics of the oscillations in $n_Q(t)$, evaluated from Fourier analysis.
Due to the non-integrability of the model, the system is expected to thermalize~\cite{polkovnikovrmp}. For small $V_f$,
the CDW gap and transition temperature is small and the effective 
temperature, the system reaches, is above the equilibrium CDW transition temperature, therefore the CDW order is absent in this case,
as evidenced in Fig. \ref{densosc2d}. For larger final interaction, however, the corresponding equilibrium transition temperature increases, and becomes equal to the effective
temperature of the thermalized system, which marks the sharp rise of $\bar n_Q$. From that point on, with increasing $V_f$, the system's effective temperature is always smaller
than the equilibrium transition temperature, therefore the system exhibits long range CDW order, mimicking a thermalized state.
This scenario is further corroborated by focusing on other characteristics of the order, such as the amplitude and frequency of the oscillations in the oscillating regime.

The equal time $\bf Q$th structure factor after the quench is obtained by inserting the time-dependent variational parameters in Eq. \eqref{strucfac}.
Its   long time average, $\bar S_c({\bf Q})$ is plotted in Fig. \ref{densosc2d}. 
At the critical $V_f$, where the transition occurs, the variational wavefunction is almost metallic,
and contains enhanced CDW correlations away from this point. However, only strong interaction can profit from these correlations and drive the dynamical CDW transition.
The effective temperature after the quench is expected to increase monotonically with $|V_f-V_i|$, while the equilibrium
CDW gap grows with $V_f$. When these two energy scales become comparable, the dynamical phase transition occurs..


\paragraph{Conclusion ---}
We have studied quantum quenches for 2D spinless fermions on the
square lattice with nearest-neighbour hopping and repulsion using the variational BWF, which are
unaccessible by exact methods. Depending on the initial and final interaction
strength, the system reaches either a time independent steady state or
oscillates periodically in time.  We argue by investigating the
characteristics of the CDW order parameter and the equal time structure factor
that the former and latter behave similarly to what is expected from a thermal
state with effective temperature larger and smaller than the equilibrium CDW
transition temperature, respectively.

Our work opens up a number of interesting questions worth pursuing in the future, such as considering bosons instead of fermions,
 the effect of disorder, the influence of imperfect nesting, incorporating the spin degree of freedom, 
as well as 
the improvement of the variational method by introducing variational parameters for each $\bf k$ mode.

\begin{acknowledgments}
We thank Johannes Motruk for sharing 2D DMRG data for the equilibrium case.
BD was supported by the Hungarian 
Scientific  Research Funds Nos. K101244, K105149, K108676 and by the Bolyai Program of the HAS.
BH was supported by the Turkish national agency for basic
research (TUBITAK grant no. 133F344).
\end{acknowledgments}

\bibliographystyle{apsrev}
\bibliography{wboson}

\pagebreak

\section{Supplementary material for "Quantum quench in 2D using the variational Baeriswyl wavefunction"}

\setcounter{equation}{0}
\renewcommand{\theequation}{S\arabic{equation}}

\setcounter{figure}{0}
\renewcommand{\thefigure}{S\arabic{figure}}

\begin{figure}[h!]
\centering
\psfrag{x}[t][][1][0]{$t$}
\psfrag{y1}[b][t][1][90]{$\alpha(t)$}
\psfrag{y2}[b][t][1][90]{$\eta(t)$}
\psfrag{V1}[][][1][0]{\color{red}$V_f=0.005$}
\psfrag{V2}[][][1][0]{$V_f=0.3$}
\psfrag{V3}[][][1][0]{\color{darkgreen}$V_f=0.59$}
\psfrag{V4}[][][1][0]{\color{magenta}$V_f=0.59125$}
\psfrag{V5}[][][1][0]{\color{blue}$V_f=0.9$}
\psfrag{V6}[][][1][0]{\color{gold}$V_f=1.2$}
\includegraphics[width=8.6cm,height=9cm]{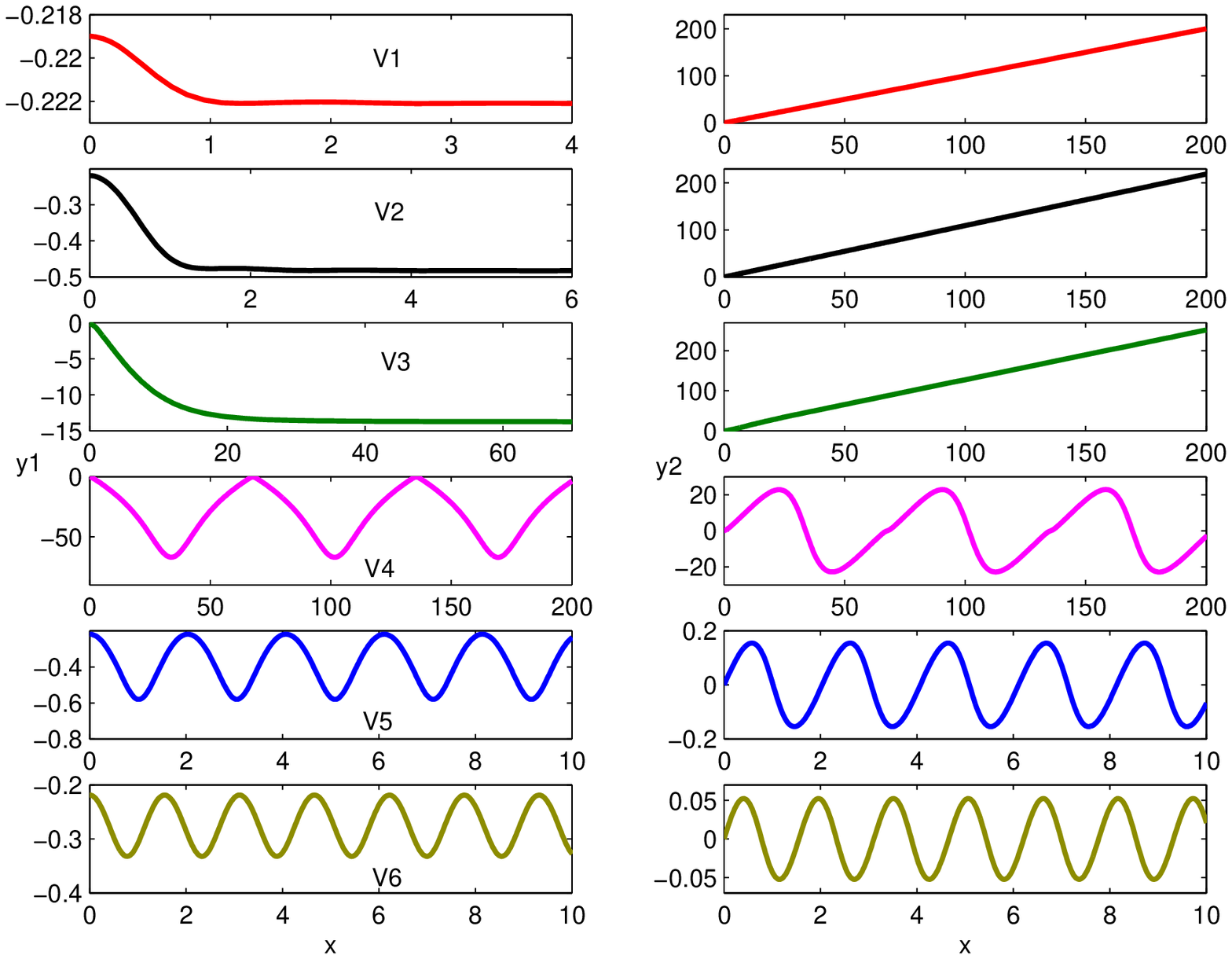}
\caption{(Color online) The time dependence of the variational parameters is plotted for $V_i=1.5$ and several final $V_f$. The critical interaction strength, separating the steady state and oscillating regimes
is around $V_f\approx 0.591$.
Note the different horizontal and vertical scales.}
\label{varpar}
\end{figure}

\section{Energy expectation value}

Due to the structure of the wavefunction, the problem is more tractable when writing the Hamiltonian $H$ in momentum space as
\begin{gather}
H_{kin}=\sum_{\bf k}\epsilon({\bf k})c^+_{\bf k}c_{\bf k},\\
 H_{int}=-\frac{V}{N}\sum_{\bf k,k^\prime,q}\epsilon({\bf q})c^+_{\bf k+q}c_{\bf k}c^+_{\bf k^\prime-q}c_{\bf k^\prime}.
\end{gather}
The interaction term can be decoupled using Wick's theorem, yielding Eq. \eqref{eint}.
The first term in Eq.~\eqref{eint} is the conventional Hartree term, $I_1$ also comes from the Hartree decoupling by taking the anomalous,
$\langle c^+_{\bf k}c_{\bf k-Q}\rangle\neq 0$ expectation values into
account. The Fock terms give rise to $I_{2,3}$, containing  normal, $\langle c^+_{\bf k}c_{\bf k}\rangle$ and anomalous expectation values, respectively.
Due to the rotational symmetry of the square lattice, the $x$ and $y$ directions are completely equivalent to each other, hence the $\cos(k_xa)$ factor
in $I_{2,3}$, coming from the $\bf q$ dependence of the interaction, together with the appropriate combinatorial factors, gives the total variational energy.
While the interaction term is even in the variational parameters, the kinetic energy is odd in $\alpha$ and independent of $\eta$, therefore their balance
determines the optimal variational parameter.

\section{Time dependence of the variational parameters}

The numerical solution of Eqs. \eqref{vareqs} yield the time-dependent variational parameters, whose behaviour is plotted in Fig. \ref{varpar}. In the steady state regime,
the $\eta(t)$ keeps on increasing linearly with time, while $\alpha(t)$ reaches a time independent steady state value. However, the relaxation time required to reach this increases with increasing
$V_f$. At a critical final interaction strength, the solution changes abruptly and both parameters exhibit periodic oscillations. The oscillation frequency increases sharply with $V_f$, and contain higher harmonics as well close to the
critical interaction strength. With increasing $V_f$, however, a single frequency describes more and more reliably the data.

\end{document}